\documentclass[aps,prb,reprint,showpacs,superscriptaddress,footinbib]{revtex4-2}

\usepackage{graphicx}
\usepackage{subfigure}
\usepackage{amsmath}
\usepackage{amsthm}
\usepackage{amssymb}
\usepackage{dcolumn}% Align table columns on decimal point
\usepackage{bm}% bold math
\usepackage{color}

\usepackage[utf8]{inputenc}
\usepackage[T1]{fontenc}
\usepackage{booktabs, array, mathptmx, float, tabularx, booktabs, lipsum, amsmath,multirow}
\usepackage{siunitx, xcolor}
\usepackage[version=4]{mhchem}
\usepackage[colorlinks,linkcolor=blue,anchorcolor=blue,citecolor=blue]{hyperref}
\newcommand{\fref}[1]{Fig. \ref{#1}}
\newcommand{\eref}[1]{Eq. \eqref{#1}}
\begin{document}

\title{Flat bands on a spherical surface from Landau levels to giant-quantum-number orbitals}

\author{Chen-Xin Jiang}
\affiliation{Department of Physics, Chongqing University, Chongqing 401331, People's Republic of China}
\affiliation{Chongqing Key Laboratory for Strongly Coupled Physics, Chongqing University, Chongqing 401331, People's Republic of China}
\affiliation{Division of Physics and Applied Physics, Nanyang Technological University, Singapore 637371}
\author{Zi-Xiang Hu\thanks{*Corresponding authors}}
\email{zxhu@cqu.edu.cn}
\affiliation{Department of Physics, Chongqing University, Chongqing 401331, People's Republic of China}
\affiliation{Chongqing Key Laboratory for Strongly Coupled Physics, Chongqing University, Chongqing 401331, People's Republic of China}
\author{Bo Yang\thanks{*Corresponding authors}}
\email{yang.bo@ntu.edu.sg}
\affiliation{Division of Physics and Applied Physics, Nanyang Technological University, Singapore 637371}

\date{\today}

\begin{abstract}
    Flat bands result in a divergent density of states and high sensitivity to interactions in physical systems.
    While such bands are well known in systems under magnetic fields,
    their realization and behavior in zero-field settings remain largely unexplored.
    Here we compare the behavior of electrons confined to a single flat band on the surface of a sphere to those in flat bands under a magnetic field.
    The zero-field flat band exhibits an additional $C(2)$ symmetry,
    which causes electrons to symmetrically cluster on opposite sides of the sphere's center when a trapping potential is introduced,
    resulting in a unique form of long-range "entanglement".
    To explore these findings experimentally,
    we propose a feasible setup to explore the unique properties of zero-field flat bands on spherical substrates,
    offering a promising route for studying interaction-driven states in spherical geometry without external fields.
\end{abstract}

\maketitle

\section{Introduction}
The flat bands in momentum space imply divergent density of states,
making the system very sensitive to interactions,
often leading to many strongly correlated phenomena.
Much of the research in this area is focused on periodic crystal structures,
such as the fractional Chern insulators \cite{articlefci,PARAMESWARAN2013816,doi:10.1142/S021797921330017X},
heavy fermion systems \cite{PhysRevLett.127.026401,PhysRevB.97.155125,PhysRevB.106.L161114},
superconductors \cite{PhysRevB.90.094506,Cao2018,PhysRevB.83.220503,doi:10.1126/science.aav1910},
Mott insulators \cite{RevModPhys.40.677,PhysRevB.101.075111}, and so on.
Research on continuous models mainly focuses on the quantum Hall effects,
where topological phases emerge from electrons confined within a two-dimensional manifold subject to a strong perpendicular magnetic field.
Here the flat bands, or the so-called Landau levels, are formed due to the magnetic field.
In the limit of large magnetic field,
only one Landau level is physically accessible at low temperature,
forming a very special Hilbert space responsible for the topological physics of both the integer and fractional quantum Hall effect.

The geometries in the quantum Hall effect mainly include the disk\cite{PhysRevLett.50.1395},
cylinder\cite{PhysRevB.107.115162}, sphere\cite{PhysRevLett.51.605}, and torus\cite{PhysRevLett.55.2095},
among which the sphere is the only curved geometry compared to the others.
The absence of boundaries on the sphere makes it particularly suitable for investigating the bulk properties of the system.
Moreover, in contrast to planar geometries, the degeneracy of Landau levels in spherical geometry is finite,
allowing for a clear definition of filling the Landau levels (LLs).

Even without a magnetic field, flat bands naturally arise on the surface of the sphere.
The single particle kinetic energies of particles confined to the spherical surface forms gapped flat bands.
Intuitively one can understand this by identifying the geometric Gaussian curvature of the sphere as analogous to a uniform "magnetic field", that nevertheless does not break time reversal symmetry.
The energy quantization in this system is governed by the electron density, the radius of the sphere, and the effective mass of the electrons.
Moreover, similar to Landau levels, the zero-field bands on the surface of the sphere remain strictly flat, with no dispersive bands present.
The study of strongly correlated systems for electrons projected onto a specific kinetic energy flat band on the surface of a sphere in the zero field remains unexplored. Previous studies look into the Coulomb interaction model of two or three electrons on the sphere without projection into a single flat band,
such as the approaches of approximate Schr\"{o}dinger equations \cite{PhysRevA.36.1575,PhysRevA.79.062517,PhysRevA.75.062506},
the configuration interaction method\cite{PhysRevA.25.1513,PhysRevA.36.511,PhysRevA.79.062517,PhysRevB.66.235118,Thompson2005ACO}. It has been suggested that electrons can form a Wigner crystal for the strong interaction limit \cite{PhysRevB.66.235118,articlecp,Thompson2005ACO,PhysRevA.75.062506,PhysRevB.97.235431}.

In this Paper, we study the physics of electrons confined to a single flat band on the surface of a conducting sphere,
as well as the mixing of flat bands from strong interactions.
Depending on the presence or absence of the magnetic field (i.e. magnetic monopole $Q$),
such flat bands have distinct physical characteristics with or without strong interaction.
Starting from a single electron,
we find that the flat band with $Q=0$ possesses an additional $C(2)$ symmetry.
This leads to the curious phenomenon that within a single flat band,
electrons cannot be localized in real space with local confining potentials.
We can thus easily create a pair of "Bell state" localized simultaneously at the North Pole and South Pole of the sphere with long-range entanglement.
Subsequently,
we study the impact of the trapping potential on the rotational symmetry of the system.
While a discrete array of trapping potentials always break the rotational invariance on the surface of the sphere,
it is not necessarily the case within a single flat band. Moreover, due to the $C(2)$ symmetry of the flat band $Q=0$,
fewer trapping potentials are needed to maintain the $\hat{L}_z$ rotational symmetry of the system,
as compared to the flat bands with $Q>0$.
We also analyze the two-body interaction within these flat bands,
and discovered that the effective interaction in $Q=0$ bands are very long-range compared to those in $Q>0$,
even from a very short-range bare interaction in the real space.
This leads to much stronger band mixing effects in the absence of the magnetic field.

\section{Results}
\subsection{Single particle flat bands}

The single-electron Hamiltonian on spherical surface with a magnetic monopole of strength $Q$ at the center of the sphere is \cite{PhysRevLett.51.605}
\begin{equation}
    \hat{H}_k=\frac{\mathbf{L}^2-\hbar^2Q^2}{2m_{\rm e}R^2},
    \label{eq:1}
\end{equation}
where $\mathbf{L}$ is orbital angular momentum, $\hbar$ is the reduced Planck constant, $m_{\rm e}$ is the effective mass of an electron and $R$ is the radius of the sphere.
We can easily get the Hamiltonian's eigenvalues $E_k^l=\frac{l(l+1)\hbar^2-\hbar^2Q^2}{2m_{\rm e}R^2}$, and the eigenstates are given by the monopole harmonics $Y_{Q,l,m}$ \cite{WU1976365,PhysRevD.16.1018},
satisfying $\mathbf{L}^2Y_{Q,l,m}=l(l+1)\hbar^2Y_{Q,l,m}$ and $\hat{L}_zY_{Q,l,m}(\bm{\Omega})=m\hbar Y_{Q,l,m}(\bm{\Omega})$,
where $l$ and $m$ denote orbital angular momentum $\mathbf{L}$ and its $z$ component $\hat{L}_z$ respectively,
and $\bm{\Omega}$ is the angular coordinates $\theta$ and $\phi$ on the sphere.
For the case of $Q > 0$,
we choose the latitudinal gauge
\begin{equation}
    \mathbf{A}=-\frac{\hbar Q}{eR}\cot\theta\hat{e}_\phi,
    \label{eq:2}
\end{equation}
and the monopole harmonics $Y_{Qlm}$ can be expressed using Wigner's $d$-function $d_{m,Q}^{(l)}(\theta)$ \cite{PhysRevD.16.1018}.
Its explicit form is given by
\begin{equation}
    Y_{Q,l,m} (\bm{\Omega})=\sqrt{\frac{2l+1}{4\mathrm{\pi}}}{\rm e}^{{\rm i}(m+Q)\phi} d_{m,Q}^{(l)}(\theta).
    \label{eq:3}
\end{equation}
The kinetic energy levels exhibit a series of flat bands with degeneracy $2l+1$.
For the Hamiltonian of zero field $Q=0$, its eigenstates are also called spherical harmonic functions $Y_{0,l,m}=Y_{l,m}$
\begin{equation}
    Y_{l,m}(\bm{\Omega})=\sqrt{\frac{2l+1}{4\mathrm{\pi}}}\sqrt{\frac{(l-m)!}{(l+m)!}}P_l^m\left[\cos(\theta)\right]{\rm e}^{{\rm i}m\phi},
    \label{eq:4}
\end{equation}
where $P_l^m(x)$ is associated Legendre polynomial.

\begin{figure}[!t]
    \centering
    \includegraphics[width=.45\textwidth]{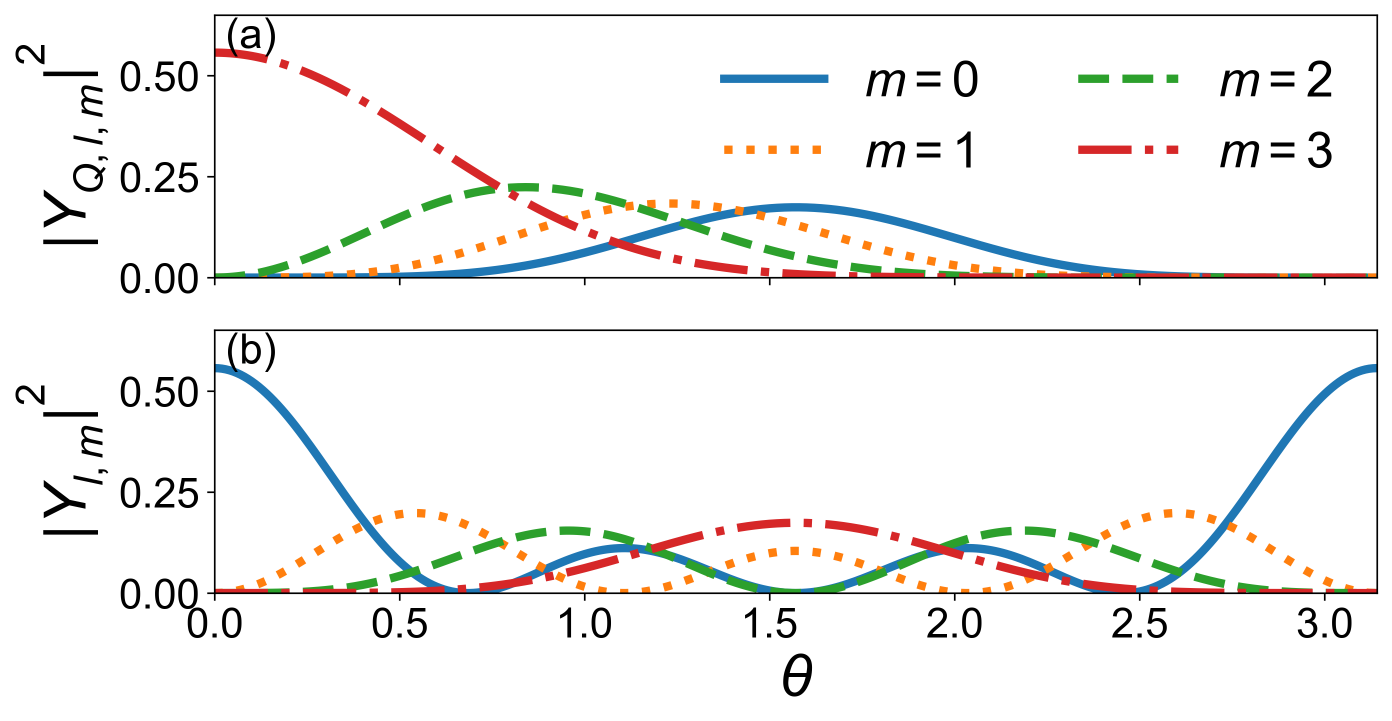}
    \caption{\textbf{Single-particle density distributions within a flat band.}
        (a) The square modulus of the monopole harmonics $Y_{Q,Q,m}$ with the strength of the magnetic monopole $Q=3$ and $\phi=0$ in the lowest Landau level, where $m$ is $z$ component of the orbital angular momentum, $\theta$ and $\phi$ are the angular coordinates.
        Blue solid, orange dotted, green dashed and red dash-dotted lines correspond to $m=0,\ 1,\ 2$ and $3$, respectively.
        (b) The square modulus of the spherical harmonic functions $Y_{l,m}$ with orbital angular momentum $l=3$ and $\phi=0$.
        Source data are provided in the Supplementary Data. }
    \label{fig:1}
\end{figure}

We focus on the dynamics within the topmost unfilled $l$th flat band where $l$ is fixed and large,
assuming that all orbitals with angular momentum less than $l$ are filled with electrons.
Consequently, as the particle density approaches infinity,
we can have a "flat band" with an infinite number of degenerate states as well as an infinitely large energy gap.
This is analogous to the Landau levels in the limit of the large system sizes (the thermodynamic limit),
where the energy gap is non-vanishing with an infinite number of degenerate states in a single Landau level.
If we fix $l$ (thus the degeneracy within the flat band),
the Landau level index is given by $l-Q$ with $Q=l$ being the Lowest Landau level (LLL).
It is well known that the eigenstates are Gaussian localized (see \fref{fig:1}(a)).
The very special $Q=0$ is the case of zero magnetic field.
From some perspectives this is the infinite Landau level limit ($l/Q\to \infty$),
though the qualitative difference is the presence of the time reversal symmetry,
so the spherical harmonics possess $C(2)$ symmetry (see \fref{fig:1}(b)).
In addition, from \eref{eq:1},
it can be concluded that the gap $\Delta E_k^l$ between the flat band with orbital angular momentum $l$ and the flat band with orbital angular momentum $l+1$ is $\frac{(l+1)\hbar^2}{m_{\rm e}R^2}$.
Unlike the equidistant distribution of Landau levels in a plane, the gap between adjacent flat bands on the sphere gradually increase as $l$ increases.
While Landau levels on the sphere have been extensively studied,
with the total flux through the surface of the sphere is an integer $2Q$ times the unit magnetic flux,
leading to $2Q + 2n + 1$ states in the $n$th Landau level (nLL),
in this work we will study the flat band at $Q=0$,
so the orbital angular momentum $l$ must be restricted to integer values.

\subsection{Localization in flat bands}

For temperature lower than the band gap,
the physical Hilbert space is a single flat band with degeneracy $2l+1$. Such flat bands are already special when we look at how we can localize a single particle within this flat band in real space.
We introduce a 1-body delta trapping potential $V^{\delta}(\bm{\Omega})$, where $\bm{\Omega}$ is the position of the electron.
We can use the Legendre expansion and the addition theorem of spherical harmonics to calculate the matrix elements of the one-body potential
\begin{equation}
    \begin{aligned}
        \mathbf{V}^{\delta,Q}_{l_1,m_1,l_2,m_2}= & \langle Q,l_1,m_1|V^{\delta}(\bm{\Omega})|Q,l_2,m_2\rangle                                                       \\
        =                                        & \sum_{l^\prime=|l_1-l_2|}^{l_1+l_2}U_{l^\prime}^{\delta}\sqrt{\frac{4\mathrm{\pi}(2l_1+1)(2l_2+1)}{2l^\prime+1}} \\
                                                 & \cdot (-1)^{Q+m_1+l^\prime+l_1+l_2} Y^*_{l^\prime,m_1-m_2}(\bm{\Omega}_0)                                        \\
                                                 & \cdot\left(\begin{array}{ccc}
                l_1 & l^\prime & l_2 \\
                -Q  & 0        & Q
            \end{array}\right)
        \left(\begin{array}{ccc}
                l_2 & l^\prime & l_1  \\
                m_2 & m_1-m_2  & -m_1
            \end{array}\right),
    \end{aligned}
    \label{eq:5}
\end{equation}
where $|Q,l,m\rangle$ denotes $Y_{Q,l,m}$,
$\bm{\Omega}_0$ is the position of the delta trapping potential in the real space on the surface of the sphere,
and the last two parentheses are wigner-3j symbols,
which can be calculated by Clebsch-Gordan coefficient \cite{Edmonds+1957}.
Furthermore, we just consider the $l$th level, so $l_1=l_2=l$. $U_{l^\prime}^{\delta}$ satisfies
$V^{\delta}(\bm{\Omega})=  \sum_{l^\prime=0}^\infty P_{l^\prime}(\cos\theta)U_{l^\prime}^{\delta}$, where
\begin{equation}
    U_{l^\prime}^{\delta}=  \frac{2l^\prime+1}{2}\int_{-1}^1V^{\delta}(\bm{\Omega})P_{l^\prime}(\cos\theta)d\cos\theta,
    \label{eq:6}
\end{equation}
where $P_l(x)$ is the Legendre polynomial,
and $\theta$ is the angle between the position of the electron and the position of the potential.
For delta potential
\begin{equation}
    V^{\delta}(\bm{\Omega})=W_\delta\delta^{(2)}\left(\left|\bm{\Omega}-\bm{\Omega}_0\right|\right),\quad U_{l^\prime}^{\delta}=W_\delta\frac{2l^\prime+1}{4\mathrm{\pi}},
    \label{eq:7}
\end{equation}
where $W_{\delta}$ is the strength of the delta potential.
To account for the effects of band mixing, the system's Hamiltonian can be expressed as
\begin{equation}
    \hat{H}_{\rm{trap}}=\hat{H}_k+\sum_{l_1,l_2,m_1,m_2}\mathbf{V}^{\delta,Q}_{l_1,m_1,l_2,m_2}C^{l_1\dagger}_{m_1}C^{l_2}_{m_2},
    \label{eq:8}
\end{equation}
where $C^{l\dagger}_{m}$ and $C^{l}_m$ denote the creation and annihilation operators for the $l$th kinetic energy level, respectively.
By diagonalizing the Hamiltonian $\hat{H}_{\rm{trap}}$,
we investigate the localization behavior of the electron within a single band and under strong band mixing conditions.

\begin{figure}[!t]
    \centering
    \includegraphics[width=.45\textwidth]{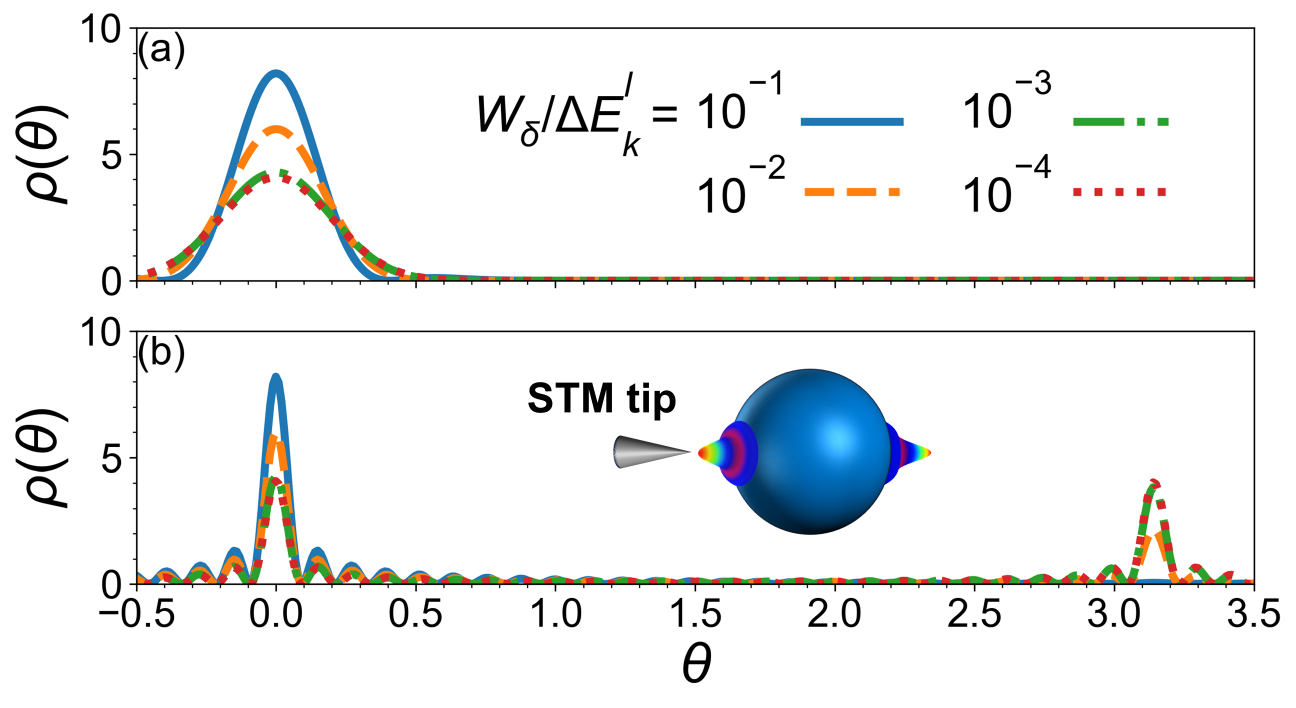}
    \caption{\textbf{Excited-state density under a delta potential.}
        The density profile $\rho$ of the excited state of a delta potential located at the North Pole for the $l$th and $(l+1)$th flat bands mixing with orbital angular momentum $l=25$.
        (a) The strength of the magnetic monopole $Q=l$ and $\phi=0$, where $\theta$ and $\phi$ are the angular coordinates.
        Blue solid, orange dotted, green dashed and red dash-dotted lines represent $W_\delta/\Delta E_k^l = 10^{-1},\ 10^{-2},\ 10^{-3}$ and $10^{-4}$, where $W_\delta$ is the strength of the delta potential and $\Delta E_k^l$ is the flat-band gap.
        (b) $Q=0$ and $\phi=0$. The gray cone marks a scanning tunneling microscopy (STM) tip applying a delta potential at one end of the blue sphere, resulting in simultaneous electron density accumulation at both ends.
        Source data are provided in the Supplementary Data.}
    \label{fig:2}
\end{figure}

In the lowest Landau level,
it is known that with a local impurity potential (e.g. in the form of a delta trapping potential),
electrons can be Gaussian localized in the real space as a coherent state as shown in \fref{fig:2}(a).
As the index of Landau level increases,
the density distribution oscillations of the excited state become more and more pronounced,
and the general expression of the real space wavefunction of the localized state at the origin $\bm{\Omega}_0=(0,0)$ in the $n$th Landau level can be written as
\begin{equation}
    \psi(\bm{\Omega})=\sum_{m=-Q-n}^{Q+n}Y^*_{Q,Q+n,m}(\bm{\Omega}_0)Y_{Q,Q+n,m}(\bm{\Omega}).
    \label{eq:9}
\end{equation}
Thus as $Q$ decreases,
it is increasingly difficult to localize a single electron within a flat band,
if the impurity potential is smaller than the band gap.

In the case of zero field with $Q=0$,
\fref{fig:2}(b) shows that the delta trapping potential produces a qualitatively different effect.
Electrons tend to cluster simultaneously at positions that are diagonally opposite on the surface of the sphere.
That is to say, within a single flat band, if we apply a delta potential at the North Pole,
we will similarly observe electron clustering at the South Pole.

When the delta potential is strong enough to mix a few flat bands,
electrons become more localized and concentrate at a single point regardless of the presence of a magnetic field,
as shown in \fref{fig:2}.
It indicates that in the case of a zero magnetic field,
band mixing breaks the $C(2)$ symmetry of the system.
This is reasonable;
for instance, if a delta potential with infinitely strong negative strength is introduced at the North Pole, the electron becomes fully localized at a single point,
which corresponds to the limit of large band mixing.
This is why it is difficult to observe this special phenomenon at $Q=0$ in reality,
as it requires the strength of the impurity potential to be much smaller than the gap $\Delta E_k^l$, as shown in \fref{fig:2}.

\subsection{Rotational symmetry in flat bands}

The special properties of the flat band Hilbert space are also reflected in its robustness of spatial symmetry. In general with localized one-body potentials or disorder, all rotational symmetries are broken.
However, in a single flat band, this issue is more subtle.
Within the flat band, the single particle $\mathbf{L}^2$ is always a good quantum number.
This is because the influence of disorder is significantly smaller than the bandwidth,
allowing the system to be regarded as a single band and preventing the mixing of different flat bands.
Consequently, the only quantum number that can be affected by the disorder is $\hat L_z$.
A single delta potential does not affect $\hat L_z$; when we introduce two such potentials, we need to place them diagonally on the sphere to ensure that $\hat{L}_z$ is still a good quantum number.
When the number of these potentials increases the rotational symmetry is generally broken, however it can be restored if the number of delta potentials is large enough and uniformly distributed, due to the unique properties of a single flat band.

For simplicity, we introduce delta potentials at equidistant points along the equator,
and ask if rotational symmetry about the $z$-axis is broken,
or if $\hat L_z$ still gives a good quantum number.
Naively this would not be the case for any number of delta potential in the full Hilbert space,
however within a single flat band the situation is more subtle.
For $Q > 0$,
when the number of delta potentials, $N_\delta$, satisfies $N_\delta \geq 2l + 1$,
$\hat{L}_z$ becomes a good quantum number.
In contrast, for $Q = 0$,
we find that $\hat{L}_z$ remains a good quantum number even when $N_\delta < 2l + 1$,
provided that $N_\delta$ is odd and greater than the orbital angular momentum $l$.
The rigorous proofs for the above phenomenon are presented in Supplementary Note 1 and 2.

This observation can also be understood intuitively as follows. Within the flat band with angular momentum $l$,
there are $2l + 1$ states, and this implies that only discrete symmetry of $C(2l)$ is present.
Consequently, if the Hamiltonian also has the same discrete symmetry,
$\hat L_z$ is restored as a good quantum number. For the flat band with zero field,
the situation is slightly different due to the additional inversion symmetry of the system.
When the number of delta potentials is odd,
it effectively corresponds to doubling the number of delta potentials in the system.
As a result, compared to systems with $Q > 0$,
the zero-field system requires fewer delta potentials to restore rotational symmetry $\hat L_z$.
In the full Hilbert space, however,
$\hat L_z$ becomes a good quantum number only in the limit where the number of delta potentials along the equator approaches infinity.

\begin{figure}[!t]
    \centering
    \includegraphics[width=.45\textwidth]{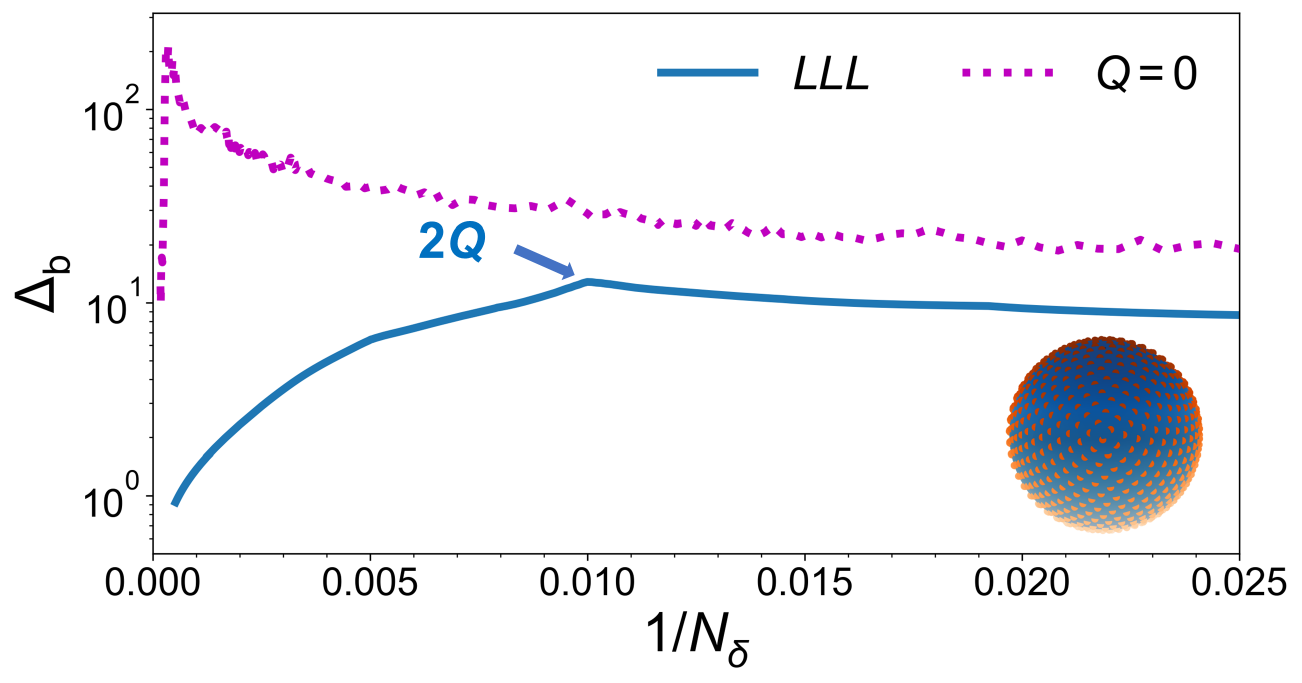}
    \caption{\textbf{Bandwidth variation and Fibonacci lattice on the sphere.}
        Energy-bandwidth as a function of $1/N_\delta$ and the number of orbitals $N_{\rm O}=101$, where the peak of the line representing the lowest Landau level (LLL) corresponds to $N_\delta = 2Q = 100$.
        $N_\delta$ is the number of delta potentials and $Q$ is the strength of the magnetic monopole.
        The blue solid line represents the LLL, while the purple dash-dotted line denotes the flat band with $Q = 0$.
        The orange dots indicate a schematic Fibonacci lattice arranged on the surface of the blue sphere.
        Source data are provided in the Supplementary Data.
    }
    \label{fig:3}
\end{figure}

When more than one electron is present,
the total angular momentum $\mathbf{L}^2=\sum_{ij}\mathbf{L}_i\mathbf{L}_j$ is generally no longer a good quantum number in the presence of impurities.
However, due to the peculiarity of the Hilbert space of a single flat band, the breaking of rotational symmetry is suppressed if we add delta potentials as evenly distributed as possible on the sphere.
We investigate the uniformity of the potential distributions within the system by analyzing the bandwidth $\Delta_{\rm b}$ of the delta potential spectrum.
Here, $\Delta_{\rm b}$ is defined as the difference between the maximum and minimum energy of the spectrum.
When the bandwidth is sufficiently narrow, this can be interpreted as delta potentials being isotropically distributed over the surface of a sphere.
While it is relatively straightforward to achieve a uniform distribution of delta potentials in toroidal geometries \cite{roy2024superlattice},
accomplishing this on a spherical surface presents significant challenges that have been extensively addressed in the literature \cite{Rakhmanov1994MinimalDE,Saff1997,articleWilliamson,González2010},
and a good solution is the Fibonacci lattice\cite{González2010},
as shown in \fref{fig:3},
where the coordinates of the $i$th point $(x_i,y_i,z_i)$ on a unit sphere are as follows
\begin{equation}
    \begin{aligned}
        z_i= & \frac{2i-1}{N_{\delta}-1},                           \\
        x_i= & \sqrt{1-z_i^2}\cos\left(2\mathrm{\pi} i \eta\right), \\
        y_i= & \sqrt{1-z_i^2}\sin\left(2\mathrm{\pi} i \eta\right),
    \end{aligned}
    \label{eq:10}
\end{equation}
where $\eta=\frac{\sqrt{5}-1}{2}$ is the golden ratio.

As shown in \fref{fig:3},
We find that when $N_{\delta}>2Q$,
the energy bandwidth $\Delta_{\rm b}$ decreases rapidly in the lowest Landau level.
This suggests that even in the presence of Fibonacci lattice delta potentials,
which break rotational invariance in the Hilbert space,
the rotational symmetry is still well approximated. This finding aligns with similar studies conducted on torus and disk geometries \cite{peng2024robusttranslationalinvariancetopological},
showing that rotational invariance is very robust against a superlattice with lattice constant smaller than the magnetic length that does not mix different LLs.
Similar to the case on the torus, the bandwidth is maximum when the number of delta potentials is equal to the number of magnetic fluxes, at which the rotational invariance is broken the most.

In contrast,
for the case of $Q = 0$,
the magnetic length no longer serves as the fundamental length scale,
although the number of states within a single flat band remains finite in analogy to the Landau levels.
Consequently, the bandwidth $\Delta_{\rm b}$ is significantly larger than in the case of $Q > 0$,
even when $N_{\delta}$ exceeds the number of states in a single flat band.
As $N_{\delta}$ becomes very large, however, $\Delta_{\rm b}$ rapidly decreases.
Nevertheless, for the same system size and the same value of $N_{\delta}$,
the rotational invariance of the $Q > 0$ band is consistently more robust,
as evidenced by its smaller bandwidth $\Delta_{\rm b}$.

\subsection{Interaction within flat bands}
We now move on to discuss the differences between the flat band with zero field and the Landau levels from the perspective of interactions. We first look at the regime when the interaction energy scale is small compared to the band gap, so that we can ignore the band mixing effect. The interaction Hamiltonian projected onto the kinetic energy level $l$ in the spherical geometry,
can be solved by projecting it into the center of mass space \cite{PhysRevLett.51.605}, and the matrix elements of the Hamiltonian can be written as
\begin{equation}
    \begin{aligned}
        \hat{H}_{\text{int}}^l= & \sum_{m_1,m_2}\sum_{m_3,m_4}\sum_{J,M}\sum_{J^\prime,M^\prime}\langle Q,l,m_1,l,m_2|Q,l,l,\mathbb{J},M\rangle      \\
                                & \cdot\langle Q,l,l,\mathbb{J},M| U(\bm{\Omega}_1-\bm{\Omega}_2)|Q,l,l,\mathbb{J}^\prime,M^\prime\rangle            \\
                                & \cdot\langle Q,l,l,\mathbb{J}^\prime,M^\prime|Q,l,m_3,l,m_4\rangle C_{m_1}^\dagger C_{m_2}^\dagger C_{m_4}C_{m_3},
    \end{aligned}
    \label{eq:11}
\end{equation}
where $U(\bm{\Omega}_1-\bm{\Omega}_2)$ is the interaction of two electrons, $\mathbb{J}$ and $M$ denote coupled total angular momentum and its
$z$ component, and $Q=0$ denotes zero field. $\langle l,m_1,l,m_2|l,l,\mathbb{J},M\rangle$ is Clebsch-Gordan coefficient \cite{Edmonds+1957},
and $\langle l,l,\mathbb{J},M| U(\bm{\Omega}_1-\bm{\Omega}_2)|l,l,\mathbb{J}^\prime,M^\prime\rangle=\mathbb{V}_{\mathbb{J}}\delta_{J,J^\prime}\delta_{M,M^\prime}$
is the 2-body pseudopotential parameter, which reflects the interaction between two electrons, and $J=2l-\mathbb{J}$ is the relative angular momentum.

\begin{figure}[!t]
    \centering
    \includegraphics[width=.45\textwidth]{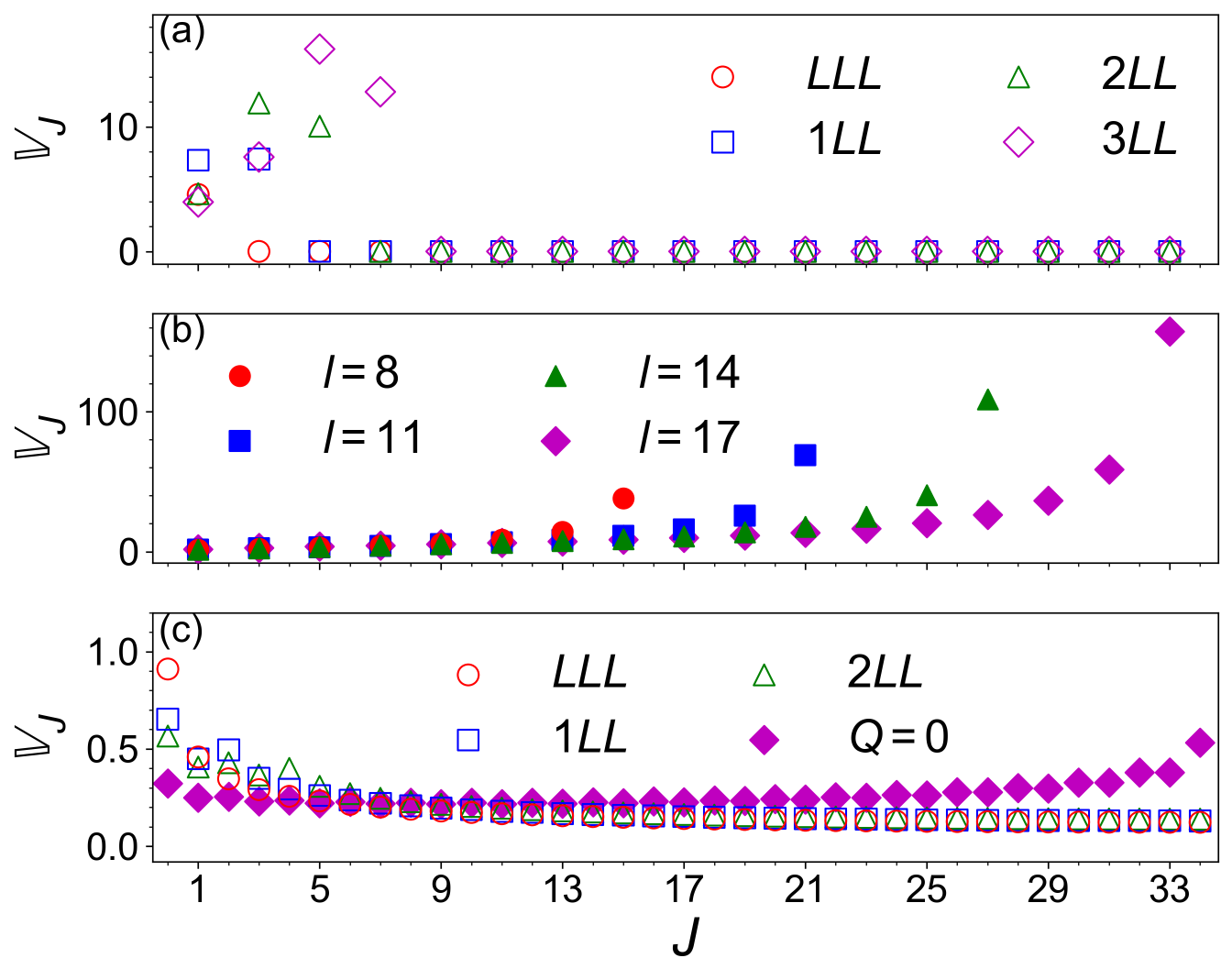}
    \caption{\textbf{Pseudopotential parameters of the Trugman-Kivelson interaction.}
        $Q$ is the strength of the magnetic monopole.
        Since for the $Q=0$ flat band, the system no longer has a magnetic length, making the sphere's radius the only relevant length scale, we perform an energy normalization by taking the radius of the sphere with zero field to be equal to that of the lowest Landau level.
        (a) Pseudopotential parameters of the Trugman-Kivelson interaction in different Landau level with orbital angular momentum $l=17$. Red open circles (LLL), blue open squares (1LL), green open triangles (2LL), and purple open diamonds (3LL) represent the lowest, first, second, and third Landau levels, respectively.
        (b) Pseudopotential parameters of the Trugman-Kivelson interaction in the $l$th flat band with $Q=0$. Red filled circles, blue filled squares, green filled triangles, and purple filled diamonds represent $l = 8$, $11$, $14$ and $17$, respectively.
        (c) Pseudopotential parameters of the Coulomb interaction in different flat band with $l=17$.
        Source data are provided in the Supplementary Data.}
    \label{fig:4}
\end{figure}

We start with the short-range Trugman-Kivelson (TK) type interaction \cite{PhysRevB.31.5280}.
For a fermionic system, it can be written as
\begin{equation}
    U^{\delta}\left(\bm{\Omega}_1-\bm{\Omega}_2\right)=\nabla^2\delta\left(\left|\bm{\Omega}_1-\bm{\Omega}_2\right|\right),
    \label{eq:12}
\end{equation}
and the corresponding pseudopotential parameters are shown in \fref{fig:4}.
For the lowest Landau level, it is well known that the 2-body pseudopotential corresponding to the TK interaction is $\mathbb{V}_1$.
This is no longer the case in higher LLs: we find that with the same contact interaction in real space, the corresponding pseudopotential is $\mathbb{V}_1+\mathbb{V}_3+\cdots+\mathbb{V}_{2l-1}$ in the $l$th Landau level.
Thus, even though different LLs are topologically equivalent, there are still qualitative differences. For example, the exact Laughlin state at $\nu=1/3$ can only be realized in the LLL.
In higher LLs the interaction tends to become more long ranged in the pseudopotential basis. This is also fundamentally related to the increase of the trace of the Fubini-Study metric in higher LLs \cite{PhysRevLett.127.246403}.

As we can understand the $Q=0$ case as the limit of an infinite number of LLs, it is not surprising that even with a very short-range interaction in the real space,
the effective interactions are very long-range in the pseudopotential basis compared to when there is a magnetic monopole,
as shown in \fref{fig:4}(b). This significant difference prevents the realization of a uniform ground state within the partially filled flat band with realistic Coulomb interactions, in contrast to the case of $Q > 0$,
as shown in Supplementary Note 3.

\subsection{Strong interaction limit and flat band mixing}
So far we only focused on the cases where the gap between the flat bands is infinite with no band mixing especially from the electron-electron interaction. In experiments the band gap is always finite, and the interaction energy scale is not necessarily small compared to the band gap. The resulting band mixing from interaction can significantly affect the dynamics of electrons. For example in the fractional quantum Hall effect,
Landau levels mixing (LLM) breaks the particle-hole symmetry of the two-body interaction system and leads to the emergence of many non-Abelian states \cite{PhysRevB.87.245425,PhysRevLett.119.026801,PhysRevLett.117.116803}.
When the Coulomb interaction is smaller than the energy gap between the flat bands,
we can treat the interaction as a perturbation to solve for an effective single Landau level model \cite{PhysRevB.74.235319,PhysRevB.87.155426,PhysRevB.42.4532,PhysRevB.98.201101}.
However, in many experimental samples, the strength of interaction is often close to or greater than the gap of flat bands \cite{PhysRevB.81.113301,PhysRevLett.108.186803},
rendering the perturbation theory ineffective.
Therefore, we choose a nonperturbative approach by directly mixing two flat bands.
Previous studies have shown that excluding higher Landau levels will not have a significant impact on the results \cite{PhysRevB.42.4532},
and the valence states are polarized \cite{PhysRevLett.80.1505,PhysRevLett.106.116801}, with excitations of different spin not substantially affecting the crucial physics of the system\cite{PhysRevLett.106.116801}.
These viewpoints can support the validity of our model.
In order to compare the different properties of the system with zero field, we mix the two flat bands with zero field and do not consider the electron spin.

The Hamiltonian of the system with flat band mixing (FBM) can be written as
\begin{equation}
    \begin{aligned}
        \hat{H}_{\rm FBM}=     & \hat{H}_k+\hat{H}_{\rm coulomb},                                                                                                                     \\
        \hat{H}_k=             & \sum_i\frac{\mathbf{L}_i^2-\hbar^2Q^2}{2m_{\rm e}R^2},                                                                                               \\
        \hat{H}_{\rm coulomb}= & \frac{e^2}{4\mathrm{\pi}\epsilon R}\sum_{m_1,m_2}\sum_{m_3,m_4}\sum_{l_1,l_2}\sum_{l_3,l_4}\langle Q,l_1,m_1,l_2,m_2|                                \\
                               & \cdot\frac{1}{\left|\bm{\Omega}_1-\bm{\Omega}_2\right|}|Q,l_3,m_3,l_4,m_4\rangle C^{l_1\dagger}_{m_1}C^{l_2\dagger}_{m_2}C^{l_4}_{m_4}C^{l_3}_{m_3},
    \end{aligned}
    \label{eq:13}
\end{equation}
where $\epsilon$ is the permittivity.
Due to the ratio of the Coulomb interaction to the gap of flat bands $\hat{H}_{\rm coulomb}/\Delta E_k^l \to R$,
we can actually control the relative strength of the Coulomb interaction by changing the radius of the sphere.

The effective Hamiltonian of the projected interaction can be written as
\begin{equation}
    \hat{H}_{\rm eff}=\sum_J\mathbb{V}^{\rm 2-body}_J\hat{\mathbb{H}}^{\rm 2-body}_J,
    \label{eq:14}
\end{equation}
which corresponds to the two-body pseudopotential parameter $\mathbb{V}^{\rm 2-body}_J$ described in \eref{eq:11},
and $\hat{\mathbb{H}}^{2{\rm -body}}_J$ is the two-body pseudopotential interaction with total relative angular momentum $J$ without FBM.
In general, the effective Hamiltonian within a single flat band that captures the band mixing effect is given by two-body or more pseudopotentials as shown below:
\begin{equation}
    \hat{H}_{\rm effm}=\sum_{i=2}\sum_{J,\alpha}\mathbb{V}^{i{\rm -bodym}}_{J,\alpha}\hat{\mathbb{H}}^{i{\rm -bodym}}_{J,\alpha},
    \label{eq:15}
\end{equation}
where $\hat{\mathbb{H}}^{i{\rm -bodym}}_{J,\alpha}$ is the $i$-body pseudopotential interaction with total relative angular momentum $J$ within a single band \cite{PhysRevB.75.195306},
$\mathbb{V}^{i{\rm -bodym}}_{J,\alpha}$denotes the corresponding pseudopotential parameters,
and $\alpha$ denotes the degeneracy of such pseudopotentials with the same total relative angular momentum.
The energy spectrum of \eref{eq:15} is defined to match the low-energy part of the full Hamiltonian in \eref{eq:13} for any number of electrons.
Consequently, the modification to the two-body pseudopotential parameter due to FBM can be easily determined as $\mathbb{V}^{\rm 2-bodym}_J-\mathbb{V}^{\rm 2-body}_J$.
Similarly, the modification to the three-body pseudopotential parameter is given by $\mathbb{V}^{3{\rm -bodym}}_{J,\alpha}$,
which can be determined by subtracting the three-electron energy contribution of the two-body effective interactions,
$\sum_J\mathbb{V}^{\rm 2-bodym}_J\hat{\mathbb{H}}^{\rm 2-bodym}_J$,
from the low-energy part of the Hamiltonian in \eref{eq:13} for three electrons.
Unlike the two-body pseudopotential coefficients,
the allowed values of $J$ for the three-body pseudopotential coefficients follow the relation $J = 3n_1 + 2n_2$ with $n_1 \geq 1,\ n_2 \geq 0,\ n_1, n_2 \in \mathbb{Z}$ \cite{PhysRevB.75.195306},
indicating that for a given $J$, multiple pseudopotential coefficients may exist.

\begin{figure}[!t]
    \centering
    \includegraphics[width=.45\textwidth]{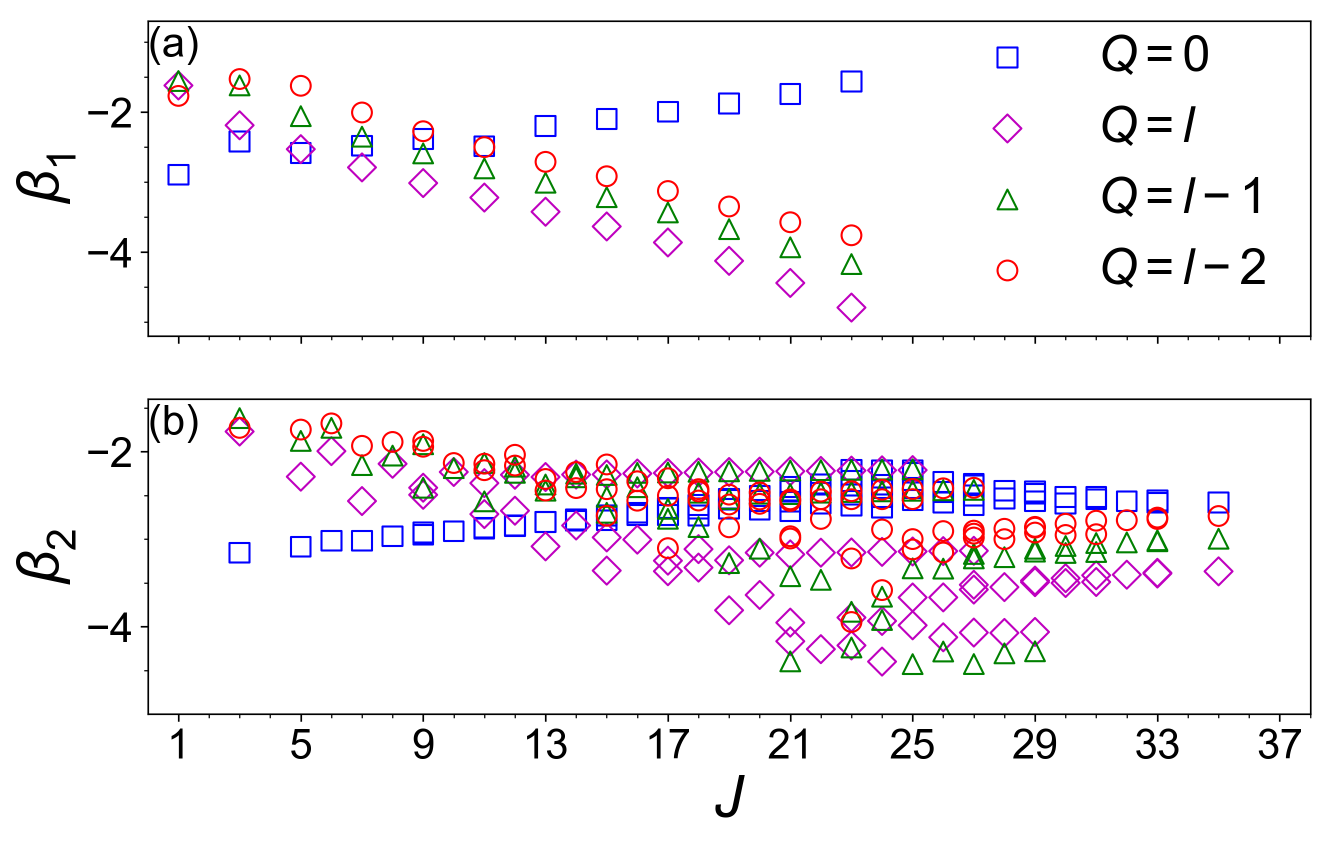}
    \caption{\textbf{Modifications to two-body and three-body pseudopotentials.}
    The effective pseudopotential parameters for the $l$th and $(l+1)$th flat bands mixing with orbital angular momentum $l=12$ and different strength of magnetic monopole $Q$, where the sphere radius for $Q=0$ is the same as that for $Q=l$. All modifications are negative values.
    (a) $\beta_1 = \lg\left|\frac{\mathbb{V}^{2{\rm -bodym}}_{J} - \mathbb{V}^{2{\rm -body}}_{J}}{\mathbb{V}^{2{\rm -body}}_{J}}\right|$ represents the logarithmic ratio of the modification to the two-body pseudopotential parameters, to the two-body pseudopotential parameters within a single flat band.
    Blue open squares, purple open diamonds, green open triangles, and Red open circles represent $Q = 0$, $l$, $l - 1$, and $l - 2$, respectively.
    (b) $\beta_2=\lg\left|\frac{\mathbb{V}^{3{\rm -bodym}}_{J,\alpha}}{\mathbb{E}_{J,\alpha}}\right|$ denotes the logarithmic ratio of the modification to the three-body pseudopotential parameters, to the three-electron energy $\mathbb{E}_{J,\alpha}$ from the two-body effective interaction $\sum_J\mathbb{V}^{\rm 2-bodym}_J\hat{\mathbb{H}}^{\rm 2-bodym}_J$.
    Source data are provided in the Supplementary Data.}
    \label{fig:5}
\end{figure}

The mixing of the $l$th and $(l+1)$th flat bands results in the pseudopotential modifications for the two-body and three-body interactions in the $l$th band, as shown in \fref{fig:5}.
As $Q$ decreases,
the modifications of band mixing on the large-$J$ components of the pseudopotential parameters become increasingly significant,
In particular, at $Q = 0$, the modifications to the large-$J$ components are the strongest and constitute the primary contribution.

\section{Discussion}

\begin{figure}[!t]
    \centering
    \includegraphics[width=.47\textwidth]{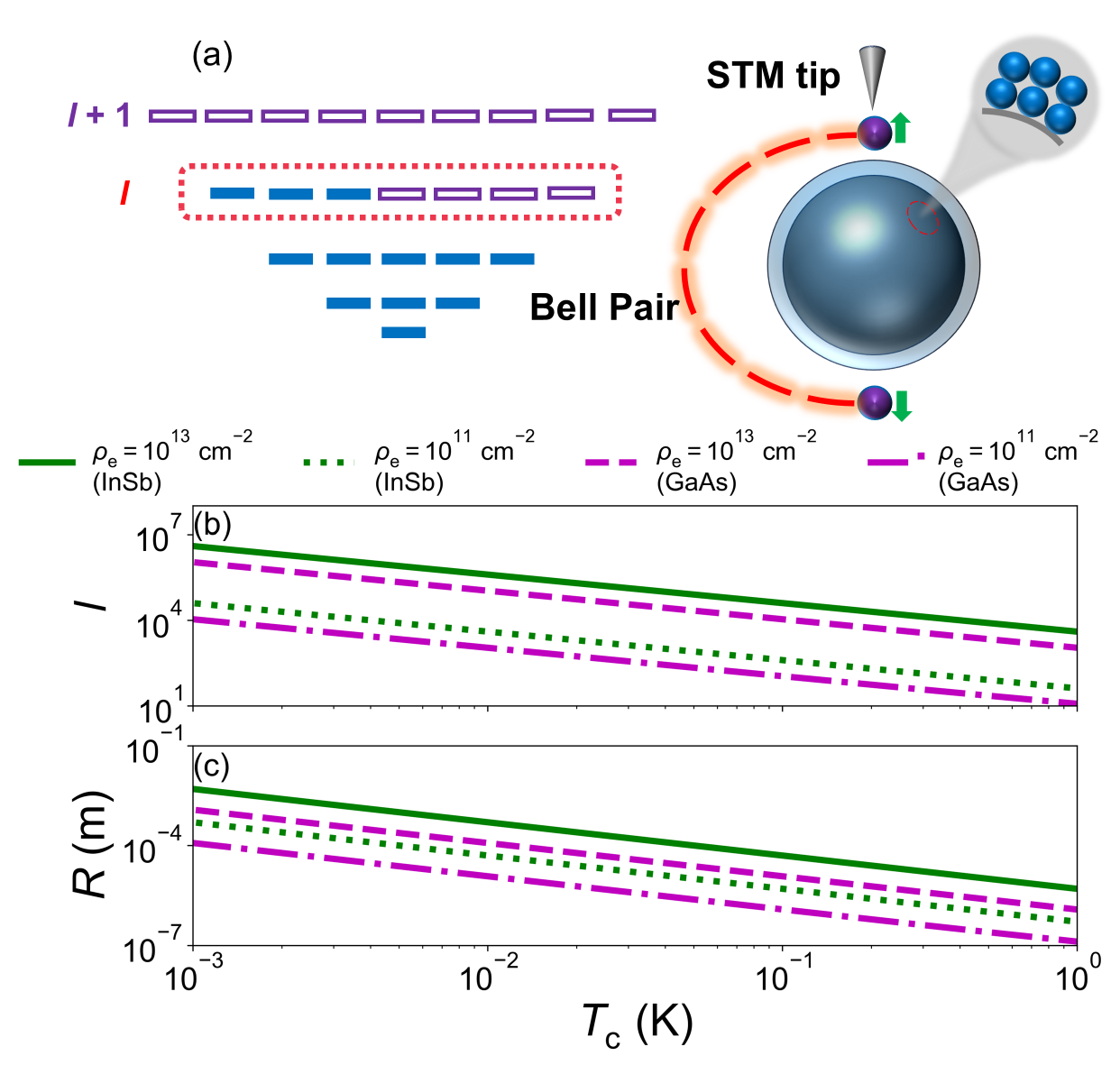}
    \caption{\textbf{Experimental concept and physical parameters for realizing spherical flat bands.}
        (a)The figure schematically illustrates the flat band on the spherical surface with the strength of the magnetic monopole $Q = 0$ and a zero-field system designed to host a Bell pair,
        where the blue solid rectangles represent orbitals occupied by electrons, the purple hollow rectangles indicate unoccupied orbitals, and the region enclosed by red dotted lines corresponds to the partially filled flat band with orbital angular momentum $l$.
        The inner sphere represents the spherical substrate,
        while the outer shell represents a layer of two-dimensional electron gas (2DEG), such as that in indium antimonide (InSb) and gallium arsenide (GaAs).
        The gray cone indicates the scanning tunneling microscopy (STM) tip applying a localized potential.
        The purple spheres at the North Pole and South Pole denote a pair of entangled electrons,
        with green arrows indicating the spin orientations of each electron.
        (b) The orbital angular momentum $l$ as functions of crossover temperature $T_{\rm c}$ for different electron densities $\rho_{\rm e}$.
        Green solid and dotted lines correspond to InSb with $\rho_{\rm e} = 10^{13}\ \mathrm{cm}^{-2}$ and $10^{11}\ \mathrm{cm}^{-2}$, respectively;
        purple dashed and dash-dotted lines correspond to GaAs with the same electron densities.
        (c) The radius $R$ as functions of crossover temperature $T_{\rm c}$ for different $\rho_{\rm e}$.
        Source data are provided in the Supplementary Data.
        \label{fig:6}}
\end{figure}

In summary, the flat band with $Q=0$ exhibits a range of intriguing properties,
one of which is its ability to host long-range entangled states with a simple experimental setup.
Specifically, as discussed in the Results section titled Localization in flat bands, a single flat band offers a rather natural setting for long-range "entanglement", especially if the sphere is sufficiently large.
A conducting spherical shell can thus be used as a convenient platform for hosting a "Bell pair", when a pair of electrons with opposite spins is added to the shell in the presence of a localization potential at the North Pole (e.g. from a scanning tunneling microscopy tip).
Due to the special $C(2)$ symmetry of the system with $Q=0$,
the probability density of the electrons is concentrated at opposite ends of the sphere even though the two ends are spatially well separated, and the local potential is only applied at one end.
The single particle long-range entanglement of the electrons of opposite spin (e.g. if they form a singlet) implies a spin-up measured at the North Pole forces a spin-down to be measured at the South Pole, or vice versa, as shown in \fref{fig:6}(a).

Several conditions need to be satisfied in the experiments for the zero-field (i.e. $Q=0$) flat band to be realized with large enough band gap as compared to temperature or disorder in the system.
Explicitly the radius $R$ of the sphere must satisfy $R\ll\sqrt{\frac{(l+1)\hbar^2}{k_{\rm B}Tm_{\rm e}}}$,
where $k_{\rm B}$ is the Boltzmann constant.
Hence, if we take $\Delta E_k^l \geq 10k_{\rm B}T$, this gives the maximum value of radius
\begin{equation}
    R_{\text{max}}=\sqrt{\frac{(l+1)\hbar^2}{10k_{\rm B}Tm_{\rm e}}}.
    \label{eq:16}
\end{equation}
Furthermore, for the flat band with orbital angular momentum $l$,
lower flat bands are completely filled, requiring a total number of $2l^2$ electrons.
The relationship between the electron density $\rho_{\rm e}$ and the radius of the sphere is thus
\begin{equation}
    R=\sqrt{\frac{l^2}{2\mathrm{\pi}\rho_{\rm e}}}.
    \label{eq:17}
\end{equation}
To determine the crossover  temperature $T_{\rm c}$,
we use the condition $R = R_{\text{max}}$,
and
\begin{equation}
    T_{\rm c}=\frac{(l+1)\mathrm{\pi}\hbar^2\rho_{\rm e}}{5l^2k_{\rm B}m_{\rm e}}.
    \label{eq:18}
\end{equation}

Realizing the flat bands at higher temperatures and on larger spherical surfaces requires materials with a sufficiently small effective mass and a high electron density,
as shown in \fref{fig:6}(b) and \fref{fig:6}(c).
If a larger spherical surface is required,
the experimental temperature must be reduced.
Alternatively, increasing the electron density provides another feasible approach.
Experimentally, several methods have been developed to achieve higher electron densities \cite{adma.202205714}.
We can look at two common materials indium antimonide (InSb) and gallium arsenide (GaAs),
with effective masses $0.014 m_0$ and $0.067 m_0$,
where $m_0$ is the free electron mass,
and electron densities $2.4\times10^{11}\ {\rm cm}^{-2}$ and $1.3\times10^{11}\ {\rm cm}^{-2}$, respectively \cite{PhysRevLett.131.266502,PhysRevResearch.4.013039,6824805}.

We also examine the effect of interaction on band mixing with realistic experimental parameters.
The scale of the Coulomb interaction is given by $\frac{e^2}{4\mathrm{\pi}\epsilon R}$.
The ratio of the interaction scale to the flat-band gap can be calculated as $\frac{m_{\rm e} e^2 R}{4\mathrm{\pi} \epsilon (l+1)\hbar^2}$.
It is well known that for $Q > 0$, $R = \sqrt{Q}l_B$, where $l_B = \sqrt{\frac{\hbar}{eB}}$ is the magnetic length.
Consequently, Landau level mixing can be suppressed by either increasing the magnetic field or decreasing $Q$.
However, in the case of $Q=0$ we no longer has a magnetic length, so the radius of the sphere is the only length scale.
For Coulomb interaction at zero field not to mix different flat bands,
it needs to be smaller than the gap $\Delta E_k^l $ of the flat band and greater than $k_{\rm B}T$,
so the radius $R$ of the sphere must satisfy two conditions.
The first condition is $R \ll \frac{4\mathrm{\pi}\epsilon(l+1)\hbar^2}{m_{\rm e} e^2}$.
Here we take
\begin{equation}
    R_1 = \frac{4\mathrm{\pi}\epsilon(l+1)\hbar^2}{10m_{\rm e} e^2},
    \label{eq:19}
\end{equation}
which ensures $R \leq R_1$.
The second condition is  $R < \frac{e^2}{4\mathrm{\pi}\epsilon k_{\rm B} T}$, and we take
\begin{equation}
    R_2 = \frac{e^2}{8\mathrm{\pi}\epsilon k_{\rm B} T},
    \label{eq:20}
\end{equation}
ensuring $R \leq R_2$.
We can use the first condition in \eref{eq:19} and \eref{eq:17} to derive the minimum value of electron density
\begin{equation}
    \rho_{\rm min}= \frac{25 m_{\rm e}^2 e^4 l^2}{8 \mathrm{\pi}^3 \epsilon^2 (l+1)^2 \hbar^4},
    \label{eq:21}
\end{equation}
For large values of $l$, the ratio $\frac{l^2}{(l+1)^2}$ approaches $1$.
This implies that to reduce the required electron density, materials with a larger permittivity and a smaller effective mass are needed, such as InSb with $\epsilon = 17 \epsilon_0$ ($\epsilon_0$ is the vacuum permittivity) \cite{PhysRevB.2.4053},
and $\rho_{\rm min} \approx 3.8 \times 10^{11} \, \text{cm}^{-2}$.
Using this density and the second condition in \eref{eq:20},
we can calculate the required radius of the sphere at a temperature of $1$ mK,
which is approximately less than $0.49$ mm.

Overall, at zero magnetic field,
a flat band can be realized on a conducting sphere with a radius on the order of $10$ to $100$ microns,
featuring an energy gap ranging from approximately $0.07$ to $0.7$ K and a degeneracy of $10^3$ to $10^4$ states.
These parameters can be further optimized with improved material properties,
such as a lower effective electron mass, higher electron density,
and larger permittivity.
In comparison, under a strong magnetic field of $10$ tesla,
a sample of approximately 1 micron by 1 micron hosts a similar order of magnitude for the number of states in each Landau level flat band,
also around $10^3$ states.
InSb emerges as a promising candidate,
capable of supporting flat bands with a degeneracy of up to $10^5$ at $1$ mK,
corresponding to a maximum sphere radius of approximately $0.78$ mm.
The fabrication of micron-scale spherical substrates is well-established across various experimental fields \cite{Okamoto2014,Wilcox_Berg_1994},
making this setup feasible. For higher temperatures or larger spheres,
increasing the electron density of the material becomes essential,
which is equally important for realizing flat bands dominated by interactions without significant band mixing effects.
It is noteworthy that the system considered here fundamentally differs from the conventional flat bands realized in quantum materials in the thermodynamic limit, where the degeneracy of the flat band approaches infinity.
In our study, we focus on a finite spherical system that, nevertheless, can achieve a high degeneracy of orbitals at the same energy (mimicking the flat bands) with a large gap from the rest of the spectrum,
potentially with experimentally accessible parameters.
Furthermore,
in conventional flat band systems,
the large degeneracy in the thermodynamic limit comes with a finite band gap,
while in our case the gap also increases with the degeneracy,
a useful property from an experimental point of view.

Future work could explore how the generation of Bell pairs on spherical systems can be experimentally realized with conducting materials with high electron density.
In the Results section titled Rotational symmetry in flat bands, we considered a highly tuned distribution of delta potentials,
highlighting the potential impact of disorder on the system.
For orbitals with large angular momentum,
the energy gap between adjacent orbitals is significantly large ($\Delta E_k^l=\frac{(l+1)\hbar^2}{m_{\rm e}R^2}$),
which strongly suppresses the mixing of different flat bands due to disorder.
While the effects of disorder in Landau levels have been extensively studied \cite{roy2024superlattice,peng2024robusttranslationalinvariancetopological},
the role of disorder in the $Q = 0$ flat band on the surface of the sphere remains an open question and will be the subject of future exploration.
Furthermore, in the absence of a magnetic field,
the impact of electron spin in this system and the novel phenomena it may induce will also be investigated in detail in future studies.

\section{Acknowledgments}
This work was supported by National Natural Science Foundation of China Grant No. 12474140 and 12347101,
Chongqing Talents: Exceptional Young Talents Project No. cstc2021ycjh-bgzxm0147 and the Fundamental Research Funds for the Central Universities Grant No. 2024CDJXY022.
C.-X. Jiang acknowledges the support of the China Scholarship Council Grant No. 202406050101. B. Yang is supported by the NTU grant for the National Research Foundation, Singapore under the NRF fellowship award (NRF-NRFF12-2020-005), and Singapore Ministry of Education (MOE) Academic Research Fund Tier 3 Grant (No. MOE-MOET32023-0003) "Quantum Geometric Advantage."

\section{Data Availability}
All numerical data generated in this study are provided in the Supplementary Information and the Supplementary Data. Source Data are available with this paper.

\section{Code Availability}
The code developed to analyze the results is available from the corresponding author upon reasonable request.

\section{Author Contributions}
C.J. performed the theoretical and numerical calculations. Z.H. and B.Y. supervised the work.

\section{Competing interests}
The authors declare no competing interests.

\section{References}
\bibliography{ref.bib}

%****figure****%

\clearpage
\newcounter{suppnote}
\renewcommand{\thesection}{Supplementary Note \arabic{suppnote}}
\let\oldsection\section
\renewcommand{\section}[1]{
    \refstepcounter{suppnote}
    \oldsection*{Supplementary Note \arabic{suppnote}: #1}
}
\renewcommand{\figurename}{Supplementary Figure}
\renewcommand{\refname}{Supplementary References}
\renewcommand{\theequation}{S\arabic{equation}}
\renewcommand{\fref}[1]{Supplementary Figure \ref{#1}}
\renewcommand{\eref}[1]{Eq. \eqref{#1}}
\setcounter{equation}{0}
\appendix

\begin{widetext}

    \title{Supplementary Information for "Flat bands on spherical surface: from Landau levels to giant-quantum-number orbitals"}

    \section{Proof of Different Landau Levels' $\hat{L}_z$ Rotational Symmetry\label{ap:1}}
    In the $n$th Landau level, the matrix elements of the Hamiltonian for $N_{\delta}$ delta potentials can be written as
    \begin{equation}
        \begin{aligned}
            H^{\delta,Q}_{m_1,m_2}= & \sum_{j=0}^{N_{\delta}-1}\mathbf{V}^{\delta,Q}_{m_1,m_2}\left(\frac{\mathrm{\pi}}{2},j\frac{2\mathrm{\pi}}{N_{\delta}}\right)                                                                                      \\
            =                       & (2Q+2n+1)\sum_{j=0}^{N_{\delta}-1}{\rm e}^{-{\rm i}(m_1-m_2)\frac{2j\mathrm{\pi}}{\tilde{N}_{\delta}}}\sum_{l^\prime=0}^{2Q+2n}(-1)^{3Q+m_1+l^\prime}U_{l^\prime}^{\delta}\sqrt{\frac{4\mathrm{\pi}}{2l^\prime+1}} \\
                                    & \cdot \left|Y^*_{l^\prime,m_1-m_2}\left(\frac{\mathrm{\pi}}{2}\right)\right|
            \left(\begin{array}{ccc}
                    Q+n & l^\prime & Q+n  \\
                    m_2 & m_1-m_2  & -m_1
                \end{array}\right)\left(\begin{array}{ccc}
                    Q+n & l^\prime & Q+n \\
                    -Q  & 0        & Q
                \end{array}\right).
        \end{aligned}
        \label{eq:a1}
    \end{equation}
    We now prove that for $N_{\delta}\geq N_{O}$, where $N_{O}=2Q+2n+1$ is the number of orbitals, namely, $N_{\delta}>2Q+2n$, when $m_1-m_2\neq0$, $H^{\delta,Q}_{m_1,m_2}=0$.
    \begin{proof}
        Clearly, the two summation parts in \eref{eq:a1} are independent of each other,
        so we can just discuss the first summation part.
        We know that $-2Q-2n\leq m_1 - m_2 \leq 2Q+2n$,
        and for $N_{\delta}>2Q+2n$,
        we have $-2 < \frac{2(m_1 - m_2)}{N_{\delta}} < 2$.
        We can express the first summation part as a problem of summing the roots of a high-order equation
        \begin{equation}
            x^{N_{\delta}}-1=  0\quad\to\quad x_j=               {\rm e}^{-{\rm i}j\frac{2(m_1-m_2)}{N_{\delta}}\mathrm{\pi}}, \quad j=0,1,\dots,N_{\delta}-1,
            \label{eq:a2}
        \end{equation}
        where $\frac{2|m_1-m_2|}{N_{\delta}}<2$, so the situation where all roots are equal to 1 will not occur.
        Therefore, by Vieta's theorem, we can conclude that the first summation of \eref{eq:a1} is equal to $0$, namely, $H^{\delta,Q}_{m_1,m_2}=0\ (m_1\neq m_2)$.
    \end{proof}
    In conclusion, when $N_{\delta} \geq N_{O}$,
    the Hamiltonian $\hat{H}^{\delta,Q}$ of $N_{\delta}$ delta potentials is diagonal,
    and $\hat{L}_z$ is also diagonal.
    Therefore, $\hat{H}^{\delta,Q}$ commutes with $\hat{L}_z$, that is, $\hat{L}_z$ of system is a good quantum number.
    It is worth noting that the conclusion we have proven does not depend on $\theta$.
    In other words, for any $\theta$, If delta potentials are added to the surface of the sphere up to the number of orbitals $N_{\rm O}$, the $\hat{L}_z$ rotational symmetry of the system will be conserved.
    We also note that in fact the first summation of \eref{eq:a1} is independent of $Q$, so this conclusion also applies to the case with zero field.

    \section{Proof of Different Flat Bands' $\hat{L}_z$ Rotational Symmetry with Zero Field\label{ap:2}}
    Due to the $C(2)$ symmetry of the flat band with zero field,
    we redefine the number of delta potentials $N_{\delta}$
    \begin{equation}
        \tilde{N}_{\delta}=\frac{N_{\delta}}{\alpha},\quad \left\{\begin{aligned}
            \alpha=1 & ,\ N_{\delta}\text{ is odd},  \\
            \alpha=2 & ,\ N_{\delta}\text{ is even}.
        \end{aligned}\right.
        \label{eq:b1}
    \end{equation}
    We prove a property of spherical harmonics $Y_{l,m}$ at the equator first, that is, when $l+m$ is odd, the value of the spherical harmonics at the equator $Y_{l,m}(\mathrm{\pi}/2,\phi)$ is $0$.
    \begin{proof}
        For Legendre polynomial $P_l(x)$, we can easily get
        \begin{equation}
            \begin{aligned}
                P_l(x)= & \sum_{k=0}^{\frac{l}{2}}A_{2k}x^{2k},\quad l{\text{ is even}},      \\
                P_l(x)= & \sum_{k=0}^{\frac{l-1}{2}}A_{2k+1}x^{2k+1},\quad l{\text{ is odd}},
            \end{aligned}
            \label{eq:b2}
        \end{equation}
        where $A_k$ are constant coefficients and associated Legendre polynomial at the equator can be written as
        \begin{equation}
            P_l^m(0)=\left.(-1)^m\frac{d^m}{dx^m}P_l(x)\right|_{x=0}.
            \label{eq:b3}
        \end{equation}
        Clearly, looking at the derivative in \eref{eq:b3} and the
        polynomial powers in \eref{eq:b2}, for odd $l$, when $m$ is even,
        the derivative does not yield a constant term, namely,
        $P_l^m(0)=0$; for even $l$, when $m$ is odd, the derivative does not yield a constant term,
        and in this case, $P_l^m(0)=0$. Therefore, when $l + m$ is odd,
        the associated Legendre polynomial evaluated at $x = 0$ is $0$,
        implying that the spherical harmonics are $0$ at the equator.
    \end{proof}
    Due to the properties of the Wigner-3j symbol \cite{Edmonds+1957}, in main text's Eq. (5) the minimum value of $l^\prime$ can take is $|m_1 - m_2|$, so for zero field $\mathbf{V}^{\delta,0}_{m_1,m_2}$ can be simplified to
    \begin{equation}
        \begin{aligned}
            \mathbf{V}^{\delta,0}_{m_1,m_2}= & (2l+1)\sum_{l^\prime=|m_1-m_2|}^{2l}(-1)^{m_1+l^\prime}U_{l^\prime}^{\delta}\sqrt{\frac{4\mathrm{\pi}}{2l^\prime+1}}
            Y^*_{l^\prime,m_1-m_2}\left(\bm{\Omega}_0\right)\left(\begin{array}{ccc}
                    l   & l^\prime & l    \\
                    m_2 & m_1-m_2  & -m_1
                \end{array}\right)\left(\begin{array}{ccc}
                    l & l^\prime & l \\
                    0 & 0        & 0
                \end{array}\right).
        \end{aligned}
        \label{eq:b4}
    \end{equation}
    The Hamiltonian matrix element of $\tilde{N}_{\delta}$ delta potentials is
    \begin{equation}
        \begin{aligned}
            H^{\delta,0}_{m_1,m_2}= & \sum_{j=0}^{\tilde{N}_{\delta}-1}\mathbf{V}^{\delta,0}_{m_1,m_2}\left(\frac{\mathrm{\pi}}{2},j\frac{\mathrm{\pi}}{\tilde{N}_{\delta}}\right)                                                                             \\
            =                       & (2l+1)\sum_{j=0}^{\tilde{N}_{\delta}-1}{\rm e}^{-{\rm i}(m_1-m_2)\frac{j\mathrm{\pi}}{\tilde{N}_{\delta}}}\sum_{l^\prime=|m_1-m_2|}^{2l}(-1)^{m_1+l^\prime}U_{l^\prime}^{\delta}\sqrt{\frac{4\mathrm{\pi}}{2l^\prime+1}}
            \left|Y^*_{l^\prime,m_1-m_2}\left(\frac{\mathrm{\pi}}{2}\right)\right|
            \left(\begin{array}{ccc}
                    l   & l^\prime & l    \\
                    m_2 & m_1-m_2  & -m_1
                \end{array}\right)\left(\begin{array}{ccc}
                    l & l^\prime & l \\
                    0 & 0        & 0
                \end{array}\right),
        \end{aligned}
        \label{eq:b5}
    \end{equation}
    where the last term can be written as
    \begin{equation}
        \left(\begin{array}{ccc}
                l & l^\prime & l \\
                0 & 0        & 0
            \end{array}\right)=\frac{(-1)^{l-l^\prime}}{\frac{l^\prime}{2}!\left(-\frac{l^\prime+1}{2}\right)!}\sqrt{\frac{\mathrm{\pi}\left(l-\frac{l^\prime+1}{2}\right)!\left(l+\frac{l^\prime}{2}\right)!}{2\left(l-\frac{l^\prime}{2}\right)!\left(l+\frac{l^\prime+1}{2}\right)!}},
        \label{eq:b6}
    \end{equation}
    where all terms are constants except $\left(-\frac{l^\prime+1}{2}\right)!$ because for $|m_1-m_2|\leq l^\prime\leq 2l$.
    When $l^\prime$ is odd, we set $l^\prime = 2k + 1,\ k=0,\ 1,\ 2,\dots$, and  this term can be written as $(-k-1)!,\ k=0,\ 1,\ 2,\dots$, and its value is $\infty$,
    so the value of \eref{eq:b6} is $0$, namely $H^{\delta}_{m_1,m_2}=0$; when $l^\prime$ is even, $\left(-\frac{l^\prime+1}{2}\right)!$ is constant. Above all,
    The second summation term can be summed only for terms whose $l^\prime$ is even. Now we prove that for $\tilde{N}_{\delta}>l$, when $m_1-m_2\neq0$, $H^{\delta}_{m_1,m_2}=0$.
    \begin{proof}
        Clearly, the two summation parts in \eref{eq:b5} are independent of each other,
        so we can first discuss the second summation part.
        When $m_1 - m_2$ is odd, as discussed earlier,
        the value of the spherical harmonic function $\left|Y^*_{l^\prime,m_1-m_2}\left(\frac{\mathrm{\pi}}{2}\right)\right|$ is $0$ when $l^\prime$ is even.
        Since the second summation term can be summed only for terms whose $l^\prime$ is even,
        so the matrix element $H^{\delta,0}_{m_1,m_2}$ is always $0$; when $m_1 - m_2$ is even, $m_1 - m_2\neq 0$,
        $l^\prime$ can only take even values,
        and the second summation part can be a non-zero constant,
        so we now consider the first summation part.
        We know that $-2l\leq m_1 - m_2 \leq 2l$,
        and for $\tilde{N}_{\delta}>l$,
        we have $-2 < (m_1 - m_2)/\tilde{N}_{\delta} < 2$.
        Since $m_1 - m_2$ is even,
        we can express the first summation part as a problem of summing the roots of a high-order equation
        \begin{equation}
            x^{\tilde{N}_{\delta}}-1=  0 \quad\to\quad   x_j=    {\rm e}^{-{\rm i}j\frac{m_1-m_2}{\tilde{N}_{\delta}}\mathrm{\pi}}, \quad j=0,1,\dots,\tilde{N}_{\delta}-1,
            \label{eq:b7}
        \end{equation}
        where $\frac{m_1-m_2}{\tilde{N}_{\delta}}<2$, so the situation where all roots are equal to $1$ will not occur.
        There, by Vieta's theorem, we can conclude that the first summation of \eref{eq:b5} is equal to $0$, namely, $H^{\delta,0}_{m_1,m_2}=0\ (m_1\neq m_2)$.
    \end{proof}
    In conclusion, when $\tilde{N}_{\delta} > l$,
    the Hamiltonian $\hat{H}^{\delta,0}$ of $\tilde{N}_{\delta}$ delta potentials is diagonal,
    and $\hat{L}_z$ is also diagonal.
    Therefore, $\hat{H}^{\delta,0}$ commutes with $\hat{L}_z$, that is, $\hat{L}_z$ of system is a good quantum number in the lowest Landau level.
    We note that the range $\tilde{N}_{\delta} > l$ actually includes the range $N_{\delta} \geq N_{O}$.

    \section{Many Particles Flat Bands\label{ap:3}}
    In the main text, under the Results section titled Interaction within flat bands,
    we discussed the differences in the interaction between the zero-field flat band and the flat band with a magnetic field,
    which will lead to completely different properties of the many-electron states in these two systems.
    Here, we consider the Coulomb interaction system with multiple electrons filling the single flat band at $Q=0$.
    Due to the presence of this special long-range interaction, the system does not exhibit a uniform ground state,
    except for the fully filled case, as shown in \fref{fig:ap1}.

    \begin{figure}[!ht]
        \centering
        \includegraphics[width=.45\textwidth]{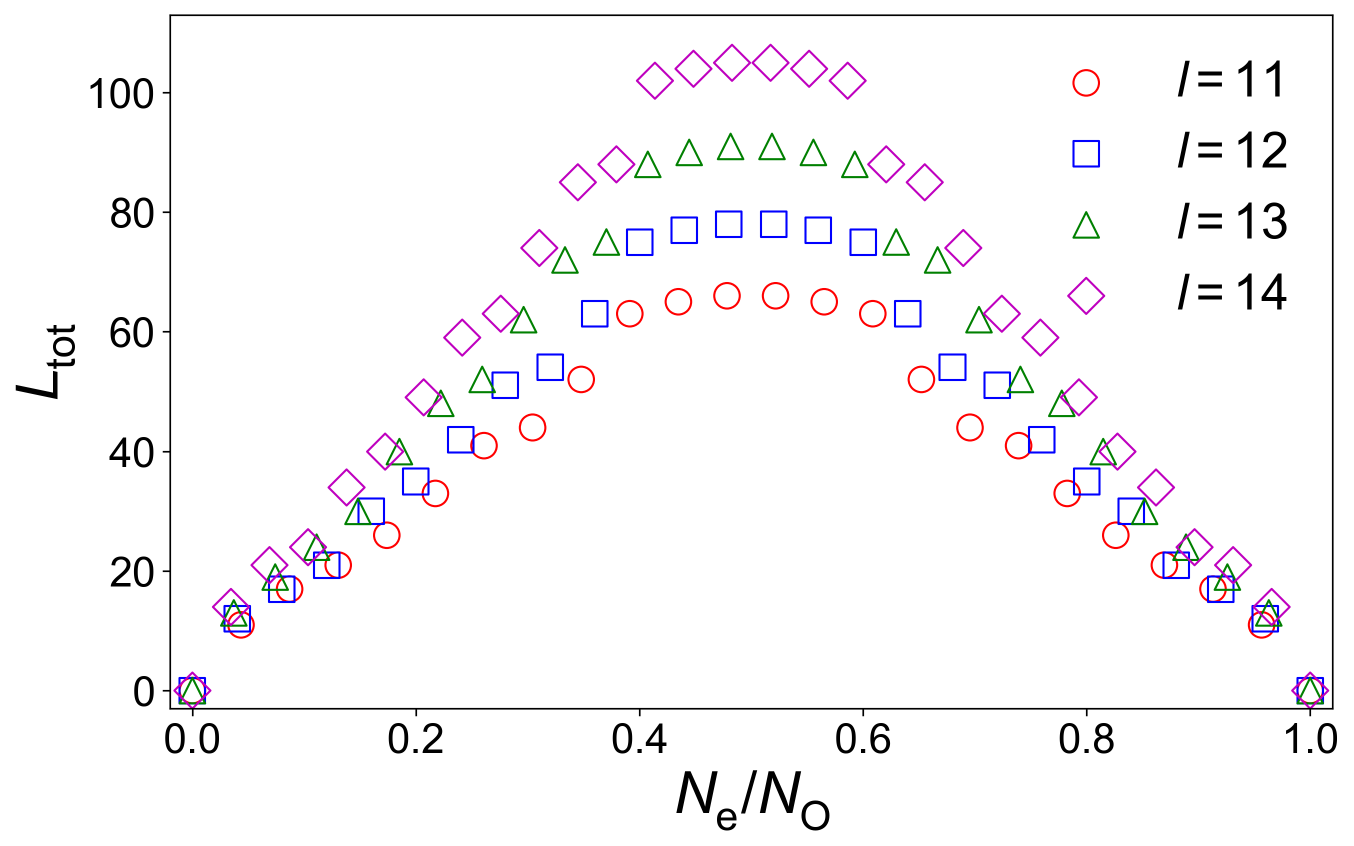}
        \caption{\textbf{Interaction-induced ground state angular momentum in a zero-field flat band.}
        Total angular momentum $L_{\rm tot}$ of the ground state with different filling $N_{\rm e}/N_{\rm O}$ in single flat band of $Q=0$,
        where $N_{\rm e}$ is the number of electrons, $N_{\rm O}$ is the number of orbitals, and $Q$ is the strength of the magnetic monopole.
        Red open circles, blue open squares, green open triangles, and purple open diamonds represent orbital angular momentum $l = 11$, $12$, $13$ and $14$, respectively.
        Source data are provided in the Supplementary Data.}
        \label{fig:ap1}
    \end{figure}

\end{widetext}

\end{document}